\documentclass[10pt,aps,prd,superscriptaddress,nofootinbib,nobibnotes,longbibliography,floatfix,twocolumn]{revtex4-2}

\usepackage{bm}
\usepackage{mathtools,
amsmath,
amssymb,
amsfonts,
mathrsfs,
chngcntr,
multirow}

\let\cc\corresponds
\let\corresponds\relax
\usepackage{mathabx}
\let\corresponds\cc

\usepackage[utf8]{inputenc}
\usepackage[T1]{fontenc}

\usepackage{soul}

\usepackage[dvipsnames]{xcolor}
\usepackage[unicode]{hyperref}
\hypersetup{colorlinks=true, citecolor=MidnightBlue,
            linkcolor=MidnightBlue, urlcolor=MidnightBlue, linktocpage=true}
\usepackage[normalem]{ulem}

\usepackage{graphicx}

\newcommand{\dV}{{\rm d}^{4}x \, \sqrt{-g}}
\newcommand{\GB}{\mathcal{G}}

\begin{document}

\title{Supermassive black hole scalarization and effective field theory}

\author{Farid Thaalba}
\email{Farid.Thaalba@nottingham.ac.uk}
\affiliation{Nottingham Centre of Gravity \& School of Mathematical Sciences, University of Nottingham, University Park, Nottingham NG7 2RD, United Kingdom}

\author{Pedro G. S. Fernandes}
\email{fernandes@thphys.uni-heidelberg.de}
\affiliation{Institut f\"ur Theoretische Physik, Universit\"at Heidelberg, Philosophenweg 12, 69120 Heidelberg, Germany}

\author{Thomas P. Sotiriou}
\email{Thomas.Sotiriou@nottingham.ac.uk}
\affiliation{Nottingham Centre of Gravity \& School of Mathematical Sciences, University of Nottingham, University Park, Nottingham NG7 2RD, United Kingdom}
\affiliation{School of Physics and Astronomy, University of Nottingham, University Park, Nottingham NG7 2RD, United Kingdom}

\begin{abstract}
A model in which black hole scalarization occurs for supermassive black holes, while their less massive counterparts remain unscalarized, has been recently proposed. We explore whether this model can emerge from an effective field theory obtained by integrating out a heavy second scalar field. We show that the resulting EFT does not have the right coupling sign or the right hierarchy of scales. We then consider whether supermassive black hole scalarization could occur in theories with two scalars. We show that, although they can violate black hole uniqueness through curvature- and spin-induced scalarization, they do not naturally produce scalarization exclusively for supermassive black holes.
\end{abstract}

\maketitle

\section{Introduction}
 Uniqueness theorems for black holes~\cite{Carter:1971zc, Chrusciel:2012jk}
 assert that all vacuum, asymptotically-flat black holes in General Relativity (GR) are described by a two-parameter family of solutions characterized by their mass and spin, known as the Kerr metric~\cite{Kerr:1963ud}. Wheeler coined the expression ``a black hole has no hair'' based on these assertions. Many no-hair theorems (or conjectures) attempt to generalize the uniqueness theorems in the presence of additional forms of matter~\cite{Chase:1970omy, Hawking:1972qk,Bekenstein:1995un,Hartle:1971qq, Teitelboim:1972ps,Bekenstein:1971hc, Bekenstein:1972ky,Sotiriou:2011dz,Sotiriou:2015pka,Herdeiro:2022yle}. 
The advent of gravitational wave astronomy, together with very-long-baseline-interferometry observations, present a great opportunity to test whether astrophysical black holes are indeed described by the Kerr metric and potentially challenge the principles of GR~\cite{LIGOScientific:2021sio, Berti:2015itd, Yunes:2024lzm, LISA:2024hlh, Gair:2012nm, Yagi:2013du, Barausse:2020rsu, LISA:2022kgy, Barack:2018yly}. In this context, deviations from the Kerr metric would indicate a breakdown in our understanding of gravity, which together with the many shortcomings of GR, and puzzles in the standard model of particle physics~\cite{Weinberg:1977ma, Damour:1994zq, Arvanitaki:2009fg, Barack:2018yly, Copeland:2006wr, Essig:2013lka, Hui:2016ltb} motivate the study of theories beyond GR and their black hole solutions.

Scalar fields are ubiquitous in extensions of GR and of the standard model.
When non-minimally coupled to gravity, scalar fields can endow a theory with non-Kerr black holes as solutions. A prominent example are scalar-Gauss-Bonnet theories~\cite{Kanti:1995vq,Yunes:2011we,Sotiriou:2013qea, Sotiriou:2014pfa, Silva:2017uqg, Doneva:2017bvd, Dima:2020yac, Herdeiro:2020wei, Berti:2020kgk, Doneva:2022ewd, Fernandes:2024ztk} where a scalar field is coupled to the Gauss-Bonnet (GB) invariant $\GB$. These theories belong to the Horndeski class~\cite{Horndeski:1974wa, Deffayet:2009mn}. 
If the coupling between the scalar field and the GB invariant is linear at leading order, then the theory admits only \textit{hairy} black holes which differ from the Kerr metric~\cite{Sotiriou:2013qea, Sotiriou:2014pfa}.
However, if the coupling is at leading order at least quadratic in a Taylor expansion around zero scalar, then the theory allows for both solutions of GR and black holes with scalar hair if they meet certain conditions; either if the black hole is sufficiently compact~\cite{Silva:2017uqg, Doneva:2017bvd}, or if it rotates fast enough~\cite{Dima:2020yac, Herdeiro:2020wei, Berti:2020kgk, Fernandes:2024ztk} (see~\cite{Doneva:2022ewd} for a review).

One way to understand how both solutions can coexist is to consider the following. For quadratic scalar-GB interactions, the linearized perturbations of the scalar $\delta \phi$ on a fixed GR background, which is a solution to the equations of motion of the theory, acquire an effective mass
\begin{align}
    (\Box - \mu^2_{\text{eff}})\delta \phi = 0,
    \label{eq:perturbations}
\end{align}
where $\mu^2_{\text{eff}}$ depends on the GB scalar and thus is spacetime dependent. As a consequence, the fluctuations might acquire a negative $\mu^2_{\text{eff}}$ in certain regions, leading to a tachyonic instability. If nonlinear interactions quench the instability, it is possible to find a new equilibrium configuration for the scalar field and the geometry --- a \textit{scalarized} configuration.

The scalar charge of a hairy black hole is fixed by its mass, spin, and the coupling constant(s) of the theory~\cite{Kanti:1995vq,Yunes:2011we,Sotiriou:2013qea,Sotiriou:2014pfa,Saravani:2019xwx, Thaalba:2022bnt}. This has been used in the literature to argue that the scalar charge will become entirely negligible for black holes that are much larger than the characteristic scale of the coupling constant(s) that control the deviations from GR or the Standard model~\cite{Maselli:2020zgv}. Constraints coming from solar mass black hole observations by the LIGO-Virgo-KAGRA collaboration~\cite{LIGOScientific:2016aoc, LIGOScientific:2017vwq, LIGOScientific:2018mvr, LIGOScientific:2020ibl}, and binary pulsars~\cite{Damour:1996ke, Yunes:2013dva, Seymour:2018bce} generically impose the length scale of new couplings to be of the order of kilometers. This, combined with the charge-mass scaling, has been used to drastically simplify the modeling of very asymmetric binaries~\cite{Maselli:2021men, Barsanti:2022vvl,Speri:2024qak} and of black hole ringdowns for massive black holes~\cite{DAddario:2023erc}.  This same fact, however, rules out the possibility that supermassive black holes could have hair in any known model, if their solar-mass counterparts do not.

Recently, Ref.~\cite{Eichhorn:2023iab} proposed a model that circumvents this argument.
In this model, solar-mass compact objects are well-described by GR solutions, while black hole uniqueness is broken at supermassive scales. The crux of this proposal is to have scalar perturbations satisfy an equation of the form~\eqref{eq:perturbations} where $\mu_{\rm eff}^2$ is only sufficiently negative as to lead to scalarization at supermassive scales, whereas at other mass/curvature scales it is positive or not negative enough to cause an instability.
The proposed model does require an auxiliary scalar field, whose dynamics are not completely clear. We will elaborate on the details in Sec.~\ref{sec:key_idea}.

Our main goal here is to examine whether we can recover the theory of Ref.~\cite{Eichhorn:2023iab} from an EFT perspective in the limit where the second scalar is heavy.
We start in Sec.~\ref{sec:scalarized_bh} by reviewing the scalarization mechanism in the standard (single-)scalar-GB theory. Then, in Sec.~\ref{sec:key_idea}, we present the model of Ref.~\cite{Eichhorn:2023iab} and discuss how the scalarization of supermassive black holes can be achieved without conflicting with current observations. In Sec.~\ref{sec:phipsi_model} we consider two interacting scalars that couple to the GB term, where one of the scalars is massive, and try to derive the model of Ref.~\cite{Eichhorn:2023iab} by integrating out the heavy scalar. We find two main drawbacks to this approach. First, integrating out the heavy field reproduces the interactions of Ref.~\cite{Eichhorn:2023iab} but with a sign difference that prevents scalarization of supermassive black holes. Secondly, the scale of the heavy field highly suppresses the novel contribution to the effective mass of scalar perturbations. In Sec.~\ref{sec:solutions}, we drop the assumption that the second scalar is heavy and study scalarization in the bi-scalar-GB model in full generality. In Sec.~\ref{sec:higgs} we show that we can obtain a coupling with the correct sign for supermassive black hole scalarization if we use a Higgs-like heavy field. However, in this case, GR vacuum geometries are no longer solutions to the theory.
We conclude in Sec.~\ref{sec:conclusions}.

\section{Theoretical Background}
\label{sec:scalarized_bh}
\subsection{Scalarization Mechanism}
Scalarization, first introduced in Ref.~\cite{Damour:1993hw} in the context of neutron stars, is a mechanism by which compact objects can be dressed up with a non-trivial scalar configuration. At the perturbative level, spontaneous scalarization manifests as a tachyonic instability that drives an exponential growth of the scalar field. For this instability to appear, scalar perturbations have to acquire a negative effective squared mass on a fixed background that is a solution of GR. This only happens if the mass or spin of the GR solution is within some range. As the scalar field grows due to this instability, perturbative analysis will eventually cease to be valid, and nonlinear effects will take control. If they drive the configuration to a new stable equilibrium, then this will be a ``scalarized'' compact object. See~\cite{Doneva:2022ewd} for a more comprehensive introduction.

Consider, for example, a theory in which the scalar field couples to the Ricci scalar as $\phi^2 R$. The effective mass of the scalar field will be proportional to the Ricci scalar, so if the Ricci scalar acquires the right sign and size in a region of spacetime, the scalar field will become tachyonically unstable. Such an effect was first illustrated in the case of neutron stars~\cite{Damour:1993hw} in a model that is equivalent to having $\phi^2R$ interaction at linear level~\cite{Andreou:2019ikc}.
Since vacuum black hole spacetimes in GR have zero Ricci scalar, scalar perturbations around them are massless in this model, and black hole scalarization does not occur. Indeed, this class of theories is covered by no-hair theorems~\cite{Hawking:1972qk, Sotiriou:2011dz}.

Black hole scalarization can be implemented in the framework of Horndeski gravity through a coupling between the scalar and the Gauss-Bonnet invariant~\cite{Silva:2017uqg, Doneva:2017bvd}
\begin{align}
    S = \int \dV \left(R + 2X + f(\phi) \mathcal{G}\right),
\end{align}
where $X=-\frac{1}{2}\partial_{\mu}\phi\partial^{\mu}\phi$ is the kinetic term and the Gauss-Bonnet invariant is given by $\mathcal{G}=R^{\mu \nu \rho \sigma} R_{\mu \nu \rho \sigma}-4 R^{\mu \nu} R_{\mu \nu}+R^{2}$, where $R_{\mu \nu \rho \sigma}$, $R_{\mu \nu}$, and $R$  being the Riemann tensor, Ricci tensor, and Ricci scalar respectively. The function $f(\phi)$ is a general function of the scalar field.
The scalar field equation is given by
\begin{align}
    \label{eq:scalar_fGB}
    \Box \phi + \frac{1}{2}f'(\phi)\mathcal{G} = 0, \quad \Box \coloneqq \nabla_{\mu}\nabla^{\mu}.
\end{align}
If $f$ is taken to be linear in $\phi$, the only allowed solutions are those with nontrivial scalar profile~\cite{Sotiriou:2013qea, Sotiriou:2014pfa}, with the linear coupling not contributing to the effective mass. If instead $f'(\phi_0)=0$ and $f''(\phi_0)\mathcal{G}<0$ for some $\phi_0$, it was shown in~\cite{Silva:2017uqg} that stationary and asymptotically flat GR black holes are unique. The first condition specifies that GR black holes are allowable, whereas the subsequent condition pertains to the mass of the scalar perturbation near $\phi_0$. Consider a perturbation $\delta \phi$ of the scalar around $\phi_0$ i.e., take $\phi = \phi_0 + \delta \phi$ on a fixed background, then Eq.~\eqref{eq:scalar_fGB} yields,    
\begin{align}
    (\Box - \mu^2_{\text{eff}})\delta \phi = 0, \quad \mu^2_{\text{eff}} = - \frac{1}{2} f''(\phi_0)\mathcal{G}  
\end{align}
therefore, the term $-f''(\phi_0)\mathcal{G}$ acts as an effective mass for the perturbations. Now, assuming that $f'(\phi_0)=0$ holds, if $f''(\phi_0)\mathcal{G}>0$ and sufficiently large, the scalar will develop the tachyonic instability associated with scalarization. The endpoint of this instability will vary depending on which other nonlinear interactions one might choose to include in the action~\cite{Silva:2018qhn, Macedo:2019sem, Antoniou:2021zoy, Staykov:2025lfh,Eichhorn:2025aja}, which can lead to scalarized black holes with different properties. For a study of the most general interactions that contribute to the onset of scalarization, see Ref.~\cite{Andreou:2019ikc}.

It is important to note that, for a Schwarzschild black hole, the GB invariant is $\mathcal{G} = 48M^2/r^6$, where $M$ is the mass of the black hole, and $r$ is the areal radius coordinate. Hence, $\GB$ is positive definite and monotonic in $r$. This implies that less massive black holes are more prone to suffer from a tachyonic instability and that above a mass threshold, scalarization will not occur because the effective mass is not sufficiently negative. 
For the Kerr geometry, the situation is more subtle since the GB invariant is not monotonic and can even become negative close to the horizon. In this case, scalarization can occur for rapidly rotating black holes with $a/M \gtrsim 0.5$, where $a$ is the spin parameter~\cite{Dima:2020yac, Herdeiro:2020wei, Berti:2020kgk}. This is referred to as spin-induced scalarization.

\subsection{Supermassive black holes scalarization}
\label{sec:key_idea}
Ref.~\cite{Eichhorn:2023iab} has proposed a model in which supermassive black holes can scalarize, while less massive ones do not. The key idea is to ensure that the effective mass of scalar perturbations is composed of various terms that compete between themselves depending on the considered curvature scale. The model, in geometric units, is 
\begin{equation}
\begin{aligned}
    S =
    \int \dV \,
    \bigg[ &R - (\partial \phi)^2 + \alpha_1 F\left(\phi\right) \GB\\& - 2 \alpha_2^3 F\left(\phi\right)
    \left( \psi\GB - \frac{\psi^2}{2} \right) \bigg],
	\label{eq:actionST}
\end{aligned}
\end{equation}
here, $\phi$ and $\psi$ are real scalar fields. The scalar field $\phi$ is dimensionless while $\psi$ is of inverse length to the fourth. Hence, the coupling constants $\alpha_1$ and $\alpha_2$ have dimensions length squared. Note that the scalar $\psi$ is auxiliary, which does not, however, mean it is not dynamical. The field equations for the scalar fields are
\begin{align}
    \psi - \GB = 0,
    \label{eq:EOMpsi}
\end{align}
and
\begin{equation}
    \Box \phi = \left[ -\alpha_1\GB + 2 \alpha_2^3 \left( \psi \GB - \frac{\psi^2}{2} \right) \right] \frac{F'\left(\phi\right)}{2}.
    \label{eq:EOMphi}
\end{equation}
Combining the two, the purpose of the field $\psi$ becomes manifest as the equation for $\phi$ becomes
\begin{equation}
    \Box \phi = \frac{1}{2} F'\left(\phi\right) \left( -\alpha_1\GB + \alpha_2^3 \GB^2 \right).
    \label{eq:EOMphi_final}
\end{equation}
To have GR solutions when $\phi=\phi_0=0$, we take the coupling function $F(\phi)$ to satisfy 
\begin{equation}
    F'(0)=0, \qquad F\left(0\right) = 0.
    \label{eq:conditions}
\end{equation}
The first condition ensures that the scalar field equation is compatible with a constant scalar field Kerr metric, while the second ensures compatibility with the field equations for the metric.
Therefore, we can consider a scalar perturbation around $\phi=0$ of the form $\phi = 0 + \delta \phi$. Using Eq.~\eqref{eq:EOMphi_final} we obtain the perturbation equation~\eqref{eq:perturbations} with an effective mass given by
\begin{equation}
    \label{eq:eff_mass}
    \mu^2_{\text{eff}} = - \alpha_1 \GB + \alpha_2^3 \GB^2,
\end{equation}
where we assumed, without loss of generality, that $F''\left(0\right)=2$. The key feature of this model becomes manifest, at least for spherically symmetric black holes: two competing terms control the effective mass of scalar perturbations. For sufficiently low curvatures, the $\GB$ term dominates and the theory reduces to the usual scalar-Gauss-Bonnet models~\cite{Silva:2017uqg, Doneva:2017bvd} discussed in previous sections. The value of $\alpha_1$ will then control what the threshold mass below which scalarization can occur, and this can be tuned to be high enough to allow supermassive black holes to scalarize.
On the other hand, if the curvature is high enough for the $\GB^2$ term to dominate, the effective mass will be positive. This suppresses scalarization for black holes with lower masses, and the value of $\alpha_2$ can be tuned to set a threshold mass for this.
Therefore, in this model, black hole uniqueness can be broken solely on supermassive black hole scales, while all other regimes agree with GR. This motivates and makes a case for experiments whose goal is to observe supermassive black holes, such as the Event Horizon Telescope~\cite{EventHorizonTelescope:2019dse,EventHorizonTelescope:2022wkp}, because deviations from GR might manifest solely in this regime. This model is also free from the early universe instability discussed in \cite{Anson:2019uto,Anson:2019ebp}. See Ref. \cite{Liu:2025mfn} for a recent discussion of the rotating black hole solutions in the model of Ref. \cite{Eichhorn:2023iab}, and Ref. \cite{Smarra:2025syw} for an analysis of how supermassive black hole scalarization affects the gravitational wave background observed by pulsar timing arrays.

In what follows, we address whether such a model can be derived from a canonical scalar field theory.

\section{Two dynamical scalar fields and effective field theory}
\label{sec:phipsi_model}
In the model of Ref.~\cite{Eichhorn:2023iab} discussed in the previous section, the dynamics of $\psi$ are unclear as there is no standard kinetic term. Here we explore if such an action can arise from a theory with a light and a heavy scalar, $\phi$ and $\psi$ respectively, after ``integrating out'' the heavy field. These fields have no relation to those in Eq. \eqref{eq:actionST}, even though we use the same symbols. We start from the following Lagrangian
\begin{align}
    \label{eq:Lagrangian_phipsi}
    \frac{2 \mathcal{L}}{M_{\text{Pl}}^2} = R - (\partial\phi)^2 - (\partial\psi)^2 - M^{2}_{\psi} \psi^2 + \alpha\phi^2\GB + 2\beta\phi\psi\GB.
\end{align}
Here, we are using units $c=1=\hbar$. The reduced Planck mass is $ M_{\text{Pl}}^{-2} = 8 \pi G$, and we have canonically normalized the scalar fields.

In this model, both scalar fields have a canonical kinetic term and are dimensionless. The coupling constants $\alpha$ and $\beta$ have dimensions of inverse mass squared (length squared). 

Action~\eqref{eq:Lagrangian_phipsi} is not the most general theory that is quadratic in the two scalars. However, we assume the $\phi$ field is light enough to ignore its mass term. A term of the form $\sim \phi \psi$ can be diagonalized away. An interaction between $\psi$ and $\GB$ of the form $\psi^2\GB$ is allowed, but its contributions would show up at higher order when integrating out the $\psi$ field. Linear interactions between the scalar fields and the Gauss-Bonnet term could be present, but they would introduce permanent hair ~\cite{Sotiriou:2014pfa,Thaalba:2022bnt}. Symmetry under  $\phi \to -\phi$ and $\psi \to -\psi$  can justify excluding these terms.  Finally, self-interactions of the scalar or couplings to $R$ could be present, but they do not affect our conclusions.

We take the scalar field $\psi$ to be ``heavy'', that is, its Compton wavelength is taken to be much smaller than that of $\phi$, and of the gravitational radius of the black hole. With this in mind, we can (formally) solve for $\psi$ using its equation of motion, and substitute it back into the Lagrangian (i.e., we are integrating out the heavy field at tree level). The equation of motion for $\psi$ is
\begin{align}
    \Box \psi - M^{2}_{\psi} \psi + \beta \phi \GB = 0 ,
\end{align}
which we can solve to leading order in $1/M^{2}_{\psi}$ as 
\begin{align}
    \psi = \frac{\beta}{M^{2}_{\psi}}\phi \GB + \mathcal{O}\left(1/M^{4}_{\psi}\right),
\end{align}
and the Lagrangian~\eqref{eq:Lagrangian_phipsi} reduces to 
\begin{align}
    \frac{2 \mathcal{L}}{M_{\text{Pl}}^2} \approx R - (\partial \phi)^2 + \alpha\phi^2\GB + \frac{\beta^2}{M^{2}_{\psi}}\phi^2\GB^2.
\end{align}
The resulting theory has the $\GB^2$ term observed in Ref.~\cite{Eichhorn:2023iab}. However, it suffers from two shortcomings. Firstly, and most obviously, the $\GB^2$ term has the ``wrong'' sign compared to theory~\eqref{eq:actionST}~\footnote{The ``correct'' sign can be obtained if $\psi$ is a tachyon.}. Secondly, the $\GB^2$ term is suppressed by the mass scale $M_\psi$. To see this, consider a Schwarzschild background 
\begin{align}
    \mathrm{d}s^2 = -\left(1-\frac{r_s}{r}\right)\mathrm{d}t^2 + \left(1-\frac{r_s}{r}\right)^{-1}\mathrm{d}r^2 + r^2 \mathrm{d}\Omega^2,
\end{align}
where $r_s$ is the Schwarzschild radius of the black hole, and let us define dimensionless coupling constants as
\begin{align}
    \Tilde{\alpha} \equiv \frac{\alpha}{r_s^2}, \quad \Tilde{\beta} \equiv \frac{\beta}{r_s^2}.
\end{align}
Then, the terms $\alpha\phi^2\GB$, and $\beta^2\phi^2\GB^2/2 M^{2}_{\psi}$ are comparable near the horizon $r\sim r_s$ if 
\begin{align}
    \label{eq:scaling}
    \tilde{\alpha} \sim \tilde{\beta}^2 \frac{1}{M_\psi^2 r_s^2},
\end{align}
Since $1 \ll M_\psi r_s$, the relation~\eqref{eq:scaling} does not hold for $\sim \mathcal{O}(1)$ dimensionless couplings $\tilde{\alpha}$, $\tilde{\beta}$. 

\subsection{Bi-Scalar Gauss-Bonnet Scalarization}
\label{sec:solutions}
In the last section, we showed that the model of Ref.~\cite{Eichhorn:2023iab} does not come from a simple effective field theory by integrating out a massive scalar. 

In this section, we will not attempt to integrate out the heavy field $\psi$. We will explore if the theory in Eq.~\eqref{eq:Lagrangian_phipsi} allows for supermassive black hole scalarization if both $\phi$ and $\psi$ are considered dynamical. 
Most of this section is dedicated to studying the tachyonic instability on a fixed Schwarzschild background and exploring the parameter space of the theory in Eq.~\eqref{eq:Lagrangian_phipsi} for which scalarized black holes are present.
The equations governing the perturbations $\delta \phi$, and $\delta \psi$ read
\begin{align}
    \Box \delta \phi &= -\alpha\GB \delta \phi - \beta \GB \delta \psi, \\
    \Box \delta \psi &= M^{2}_{\psi}\delta \psi - \beta \GB \delta \phi.
\end{align}
In a static spherically symmetric background, we can perform the following decomposition of the perturbations
\begin{align}
    \delta \Phi_{i} &= u_{i}(r) \exp(- \textrm{i} \omega t) Y_{\ell m}(\theta, \varphi) / r, \quad i \in \{1,2\}
    \label{eq:perturbations1}
\end{align}
where $\Phi_{i} = \{\phi, \psi\}$, and $Y_{\ell m}$ are the spherical harmonics. We tackle these equations for static ($\omega = 0$) monopolar ($\ell=0$) excitations in a Schwarzschild background. Note that these equations cannot, generally, be decoupled except in special cases. First, it is instructive to consider two limiting cases: (i) the large mass (``frozen'' $\psi$) case i.e., we take $\Box \delta \psi=0$, and (ii) the zero mass case $M_{\psi}=0$. Finally, we solve the system in full generality and compare it with the limiting cases.

We solve the equations in isotropic coordinates $\rho$ such that $r = \rho (1+M/2 \rho)^2$. Near the horizon $\rho_h = M/2$, we use a series expansion in $(\rho-\rho_h)$ up to second-order precision for the functions $u_{i}$ in Eq.~\eqref{eq:perturbations1} as
\begin{align}
    u_{i} = u_{i}^{(0)} + u_{i}^{(2)} \left(\rho-\frac{M}{2}\right)^2.
\end{align}
We can fix $u_{1}^{(0)}=1$ by an appropriate rescaling of the equations, and solve algebraically for $u_{i}^{(2)}$ in terms of $u_{2}^{(0)}$ and the various coupling constants, thereby establishing the boundary conditions close to the horizon. Subsequently, for a fixed value of the coupling constant $\beta$ a shooting method is utilized to determine the values of $u_{2}^{(0)}$ and $\alpha$ admitting a nontrivial solution where $u_{i}(r)$ asymptotically approach a constant (zero) value for large $\rho$ (for asymptotic flatness). We typically terminate integration at a large value $\rho=100M$ and fix $M=1$.
\begin{figure*}
    \centering
    \includegraphics[width=.47\linewidth]{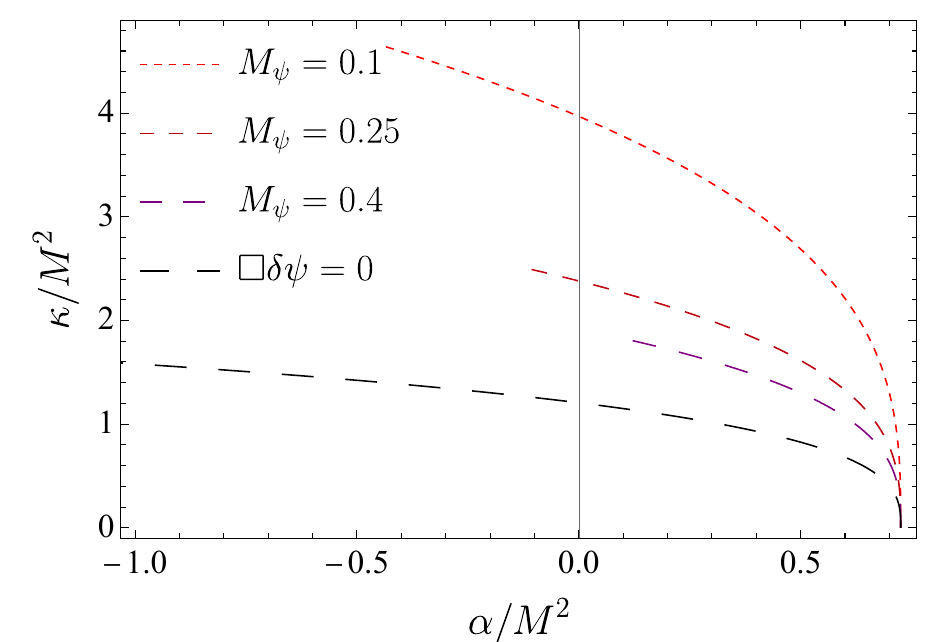}
    \hspace{5mm}
    \includegraphics[width=.47\linewidth]{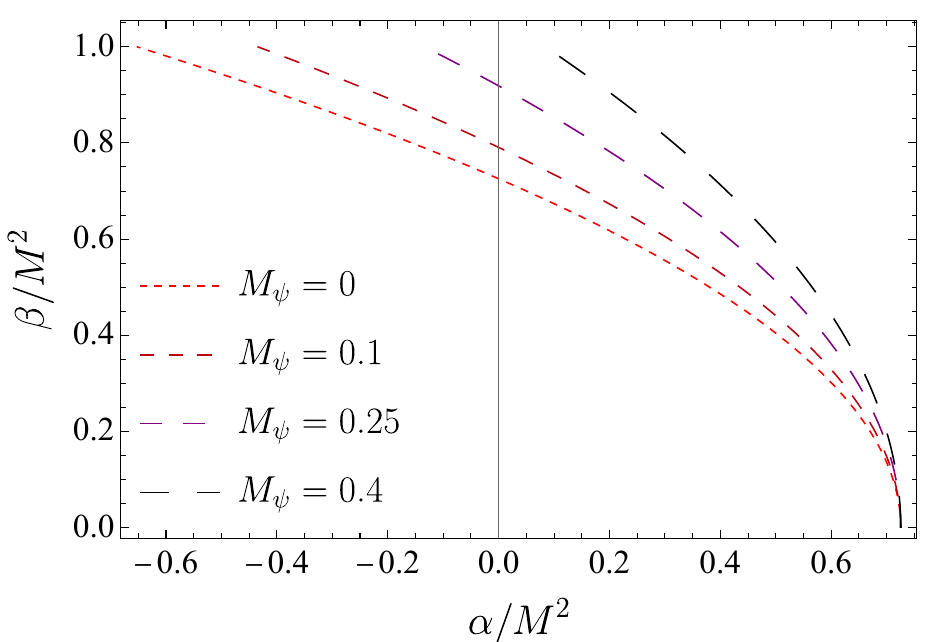}
    \caption{Parameter space of scalarized solutions for different values of $M_{\psi}$. Left: We compare the behavior of the large mass case ($\Box \delta \psi=0$) with the general case of finite non-zero $M_{\psi}$. Black holes are tachyonically unstable above and to the right of the curves. We observe that for fixed $M_{\psi}$ smaller values of $\alpha/M^2$ require larger $\kappa$ (where $\kappa^3=\beta^2/M_\psi^2$), thus larger $\beta$, to exhibit a tachyonic instability. Right: We compare the massless scenario with the more general case where $M_{\psi}$ is non-zero. As $M_\psi$ increases, for fixed $\alpha$, higher values of $\beta$ are required to allow scalarization.}
    \label{fig:par_space}
\end{figure*}

In the first limiting case, where $\psi$ is heavy, it decouples from the spectrum of the theory and we have only one scalar perturbation equation for $\delta \phi$ to solve for with an effective mass given by
\begin{align}
    \mu_{\text{eff}}^2 = - \alpha \GB - \kappa^3 \GB^2, \quad \kappa^3 \equiv \frac{\beta^2}{M^{2}_{\psi}}.
    \label{eq:2scalarsmueffheavy}
\end{align}
In the other limiting case, where we consider $\psi$ to be massless ($M_\psi=0$), the system of equations can be diagonalized. To appreciate this fact, let us rewrite the Lagrangian~\eqref{eq:Lagrangian_phipsi} in a matrix form as~\footnote{Here, we have dropped overall factors of the Lagrangian.}
\begin{align}
    \mathcal{L} = R - (\partial \Phi)^2 - \Phi \cdot \mathcal{M} \cdot \Phi,
\end{align}
where, 
\begin{align}
    \Phi = \begin{pmatrix} \phi \\ \psi \end{pmatrix}, \quad \mathcal{M} = \begin{pmatrix} -\alpha \GB & -\beta\GB \\ -\beta\GB & M^2_{\psi} \end{pmatrix}.
\end{align}
If $\psi$ is massless, then the matrix $\mathcal{M}$ can be diagonalized, with eigenvectors that are spacetime independent, and the following eigenvalues 
\begin{align}
    \lambda_{\pm} = \frac{-\alpha \pm \sqrt{\alpha^2 + 4 \beta^2}}{2}\GB, 
\end{align}
Consequently, there must exist a unitary transformation such that the Lagrangian can be written as 
\begin{align}
    \mathcal{L} = R - (\partial \phi_+)^2 - (\partial \phi_-)^2 + \alpha_{+}\phi_+^2\GB + \alpha_{-}\phi_-^2\GB,
\end{align}
for two scalar fields $\phi_+$, $\phi_-$, and with $\alpha_{+,-}$ given by
\begin{align}
    \alpha_{+} = \frac{\alpha + \sqrt{\alpha^2 + 4 \beta^2}}{2}, \quad 
    \alpha_{-} = \frac{\alpha - \sqrt{\alpha^2 + 4 \beta^2}}{2}.
\end{align}
It is important to note that $\alpha_{+}$ is always positive, and $\alpha_{-}$ is always negative, regardless of the non-vanishing values of $\alpha$ and $\beta$. This implies that in this model, both types of scalarization, curvature and spin-induced, must occur. Of course, it is trivial to write a theory with two scalar fields that allows both types of scalarization, but one has to force the couplings to be of a fixed sign in an \textit{ad-hoc} manner. However, in this model, the sign of the couplings is not free, but is instead fixed by the diagonalization process. As a consequence, as long as there is an interaction term of the form $\phi \psi \GB$ in the theory, then both types of scalarization are inevitable.

In Fig.~\ref{fig:par_space} we present the bulk of our numerical results. The results obtained for the effective mass in   Eq.~\eqref{eq:2scalarsmueffheavy}, and finite $M_\psi$ are shown in the left panel. We observe that for a fixed black hole mass $M$ and coupling $\alpha$, larger values of $\kappa$ are needed for scalarization to occur, when smaller $M_\psi$ is considered. In some cases, we observe scalarization even for negative values of $\alpha$, which does not occur when the coupling between $\phi$ and $\psi$ is absent. 
For a fixed value of $\beta$, we would need a smaller $M_{\psi}$ to acquire a negative effective mass and trigger the instability, which we observe in the left panel of Fig.~\ref{fig:par_space}.

In the right panel of Fig.~\ref{fig:par_space}, we compare the massless limit $M_\psi=0$ with the massive scalar cases. We find that, for fixed coupling $\beta$, larger values of $M_{\psi}$ demand larger values of $\alpha$ to scalarize a black hole as the mass of the scalar field will contribute positively to the effective mass of the perturbations.

We also observe that for $\beta=0$, we have $\alpha/M^2 \sim 0.726$ as the threshold for scalarization, consistent with the results of Refs.~\cite{Silva:2017uqg, Doneva:2017bvd}. This is true because for $\beta=0$ we have two decoupled scalar fields with $\phi$ coupled to gravity as in the usual scalar Gauss-Bonnet theories, while $\psi$ is just a free massive scalar field which is covered by a no-hair theorem. On the other hand, when $\alpha=0$, the bifurcation point changes depending on the value of $M_{\psi}$ as expected. 

We do not find evidence for a scenario similar to that of Ref.~\cite{Eichhorn:2023iab}, where supermassive black holes can scalarize, while solar-mass compact objects remain stable and well-described by GR.

Finally, as a remark, we note that the curves in Fig.~\ref{fig:par_space} for finite values of the scalar field mass $M_\psi$ stop at some finite value of $\alpha$. This is merely an artifact of the numerical stiffness of the perturbation equations when a finite mass $M_\psi$ is considered, because of the exponential decay of $\psi$. We expect the curves to continue on their trend to increasingly negative values of $\alpha$ as suggested by the large mass limit, where the system is not stiff anymore. The curves shown in Fig.~\ref{fig:par_space} belong to the portion of the parameter space where we were able to accurately solve the stiff system of differential equations numerically.

\section{The correct coupling sign from a Higgs-like mechanism}
\label{sec:higgs}

We now consider if a low-energy EFT that includes a mass term given by  Eq.~\eqref{eq:eff_mass}, and hence can have supermassive black hole scalarization, could arise from a Higgs-like interaction. To this effect, we start from the action
\begin{equation}
    \begin{aligned}
    \mathcal{L} =& \frac{M_{\text{Pl}}^2}{2}R - \frac{1}{2}(\partial \phi)^2 - \frac{1}{2}(\partial H)^2 - \frac{\lambda}{4} (H^2-v^2)^2 \\& - \frac{1}{2}\alpha F\left(\phi/\Lambda\right) H^2 \GB,
\end{aligned}
\end{equation}
where $H$ is a Higgs-type scalar field with vacuum expectation value (VEV) at infinity $H=v$, and $\Lambda$ is an energy scale. The main difference of this model in comparison with the one in Eq.~\eqref{eq:Lagrangian_phipsi} is precisely that the massive field has a non-zero VEV. We consider perturbations of $H$ around the VEV, $H = v + h$, for which we obtain the following Lagrangian.
\begin{align}
    \mathcal{L} &= \frac{M_{\text{Pl}}^2}{2}R - \frac{1}{2}(\partial \phi)^2 - \frac{1}{2} (\partial h)^2 \nonumber \\   
    &- \frac{\lambda}{4} (h+2v)^2 h^2 - \frac{1}{2}\alpha F\left(\phi/\Lambda\right) (h+v)^2 \GB.
\end{align}
The equation of motion for $h$ is 
\begin{equation}
    \begin{aligned}
        \Box h =& \left(\alpha F\left(\phi/\Lambda\right) \mathcal{G}  + 2\lambda  v^2\right) h +\alpha  v F\left(\phi/\Lambda\right) \mathcal{G} \\&+ 3 \lambda  v h^2+ \lambda  h^3,
    \end{aligned}
\end{equation}
where we observe that the Higgs field acquires a bare mass $m_h^2=2\lambda v^2$. Assuming the Higgs to be heavy ($m_h r_s \gg 1$), we can freeze it, $\Box h \approx 0$, and solve for $h$,
\begin{align}
    h \approx v \left(\sqrt{1-\frac{2 \alpha}{m_h^2} F\left(\phi/\Lambda\right) \mathcal{G}}-1\right),
\end{align}
and we obtain the low-energy EFT, 
\begin{align}
    \label{eq:metric_higgs}
    \mathcal{L} = \frac{M_{\text{Pl}}^2}{2}R - \frac{1}{2}(\partial \phi)^2 - \frac{\alpha v^2}{2}F\left(\phi/\Lambda\right) \mathcal{G} + \frac{v^2 \alpha^2}{2m_h^2} F\left(\phi/\Lambda\right)^2 \GB^2.
\end{align}
The equation of motion for $\phi$ obtained from the EFT is given by
\begin{align}
    \label{eq:phi_higgs}
    \Box \phi = \frac{\alpha v^2}{2\Lambda} F'(\phi/\Lambda) \mathcal{G} - \frac{\alpha^2 v^2}{m_h^2 \Lambda} F(\phi/\Lambda) F'(\phi/\Lambda) \GB^2,
\end{align}
which allows for constant scalar configurations $\phi=0$ as long as $F'(0)=0$. The scalar perturbations around the background are described by a Klein-Gordon equation with a squared effective mass given by
\begin{equation}
    \mu_{\rm eff}^2 = \frac{\alpha v^2}{2 \Lambda^2} F''(0) \GB - \frac{\alpha^2 v^2}{m_h^2 \Lambda^2} F(0)F''(0) \GB^2.
    \label{eq:pertHiggs}
\end{equation}
Using the symmetry $\alpha \to -\alpha$, together with $F\to -F$, we can fix the coupling function so that $F(0)\geq0$, ensuring the two contributions to the effective mass have opposite signs. For $F''(0)<0$ and $\alpha>0$ it is possible to reproduce the perturbation equation from Ref.~\cite{Eichhorn:2023iab}, given in Eq.~\eqref{eq:eff_mass}.

Under these circumstances ($F(0)\geq0$, $F''(0)<0$, and $\alpha>0$), there are two possible cases to consider, depending on whether $F(0)$ is vanishing or not.
\begin{enumerate}
    \item $F(0) = 0$: In this case, the Kerr metric together with $\phi=0$ solves all field equations. However, the perturbation equation reduces to the usual one of standard scalarization because the contribution containing the $\GB^2$ term vanishes.
    \item $F(0) > 0$: In this case, the Kerr metric together with $\phi=0$ no longer solves the field equations of the theory. Instead, a solution with $\phi=0$ is allowed if the metric is a solution of the theory
    \begin{equation}
        \label{eq:higgs_EFT_ctescalar}
        \mathcal{L} = \frac{M_{\text{Pl}}^2}{2}R + \frac{k}{2} \GB^2, \qquad k=\frac{\alpha^2 v^2}{m_h^2} F(0)^2.
    \end{equation}
    Then, scalar perturbations governed by Eq.~\eqref{eq:pertHiggs} are realized in this non-GR vacuum.
\end{enumerate}
Consequently, in this setting, the effective mass of scalar perturbations necessary for supermassive black hole scalarization~\eqref{eq:eff_mass} can be reproduced, but not in a theory allowing GR vacuum solutions. 

Although the results in this section suggest that an action of the type considered in Ref.~\cite{Eichhorn:2023iab} is unlikely to arise from a Higgs-like interaction in a natural way, they also suggest another way in which hair could arise for supermassive black holes only. 
Corrections to GR in the form of higher-curvature corrections are expected on general grounds.
Assuming, for instance, a coupling function which can be expanded as $F(\phi/\Lambda) = 1 - (\phi/\Lambda)^2 + \mathcal{O}\left((\phi/\Lambda)^3\right)$, the effective mass~\eqref{eq:pertHiggs} becomes
\begin{equation}
    \mu_{\rm eff}^2 = \frac{v^2}{\Lambda^2} \left(-\alpha \GB + \frac{2\alpha^2}{m_h^2}\GB^2 \right),
\end{equation}
which can be brought to the form of Eq.~\eqref{eq:eff_mass}, with $\alpha_1 = \alpha v^2/\Lambda^2$ and $\alpha_2^3 = 2v^2\alpha^2/(\Lambda^2 m_h^2)$. From the results of Ref.~\cite{Eichhorn:2023iab}, we expect scalarization of supermassive black holes to be possible if $\alpha_1 \sim \alpha_2$, which translates to $\Lambda^2/v^2\sim m_h \sqrt{\alpha}$.
Moreover, a positive contribution to the effective mass that can prevent the theory from exhibiting catastrophic instabilities in the early universe~\cite{Anson:2019uto,Anson:2019ebp}, as discussed in Ref.~\cite{Eichhorn:2023iab}, is produced.

\section{Conclusions and Discussion}
\label{sec:conclusions}
One of the central observational and theoretical questions in modern physics is whether all astrophysical dark compact objects are accurately described by the Kerr metric, and if not, in what regimes the predictions of GR break down. In most models where deviations from the Kerr metric arise, these deviations typically occur for smaller black holes, making solar-mass black holes particularly promising candidates for testing gravity. However, recently Ref.~\cite{Eichhorn:2023iab} proposed a model where the opposite occurs: black hole uniqueness is violated at supermassive scales, while solar-mass objects remain well-described by GR. This finding underscores the importance of experiments targeting different mass ranges of astrophysical compact objects, such as the Event Horizon Telescope or LISA.

The model proposed in Ref.~\cite{Eichhorn:2023iab} features two scalar fields, one of which is auxiliary but dynamically active, though its precise role and behavior remains unclear. This auxiliary field mediates a coupling between the other scalar field and a higher-order curvature term. As a result, the effective mass of scalar perturbations is determined by two competing terms proportional to $\GB$ and to $\GB^2$, with opposite signs. This competition allows for scalarization within a specific mass range, which can include supermassive black holes, offering a novel mechanism for breaking black hole uniqueness at these scales. Our goal here was to explore whether such a model can arise naturally in the context of effective field theory.

We have first investigated whether we can recover the model from a theory with two canonical scalar fields by assuming that one scalar is heavy and remains unexcited in a black hole background. Integrating out this heavy field produces an EFT description of the system, leading to an emergent coupling between the light scalar and $\GB^2$. However, our analysis uncovered two key limitations: first, the generated term acquires the ``wrong'' sign upon integrating out the heavy field, preventing it from contributing to the effective mass of perturbations in a manner that would enable scalarization of supermassive black holes; second, dimensional analysis reveals that this interaction is significantly suppressed compared to the $\GB$ interaction.

Then, we relaxed the assumption that the second scalar field is heavy enough to be integrated out and analyzed scalar perturbations on a fixed Schwarzschild background of a theory with two dynamical scalars described by Eq.~\eqref{eq:Lagrangian_phipsi}. In the limit in which both scalars are massless, we were able to diagonalize and obtain two decoupled scalar fields, each coupling quadratically to $\GB$ with coupling constants of opposite signs. Interestingly, this ensures that both curvature- and spin-induced scalarization \emph{must} occur within the same theory. In the more general case, we solved the full system and explored the theory's parameter space. Our analysis revealed no evidence of a scalarization mechanism capable of breaking black hole uniqueness at supermassive scales.  

Next, we examined a similar model in which the heavy scalar is a Higgs-like field. Upon integrating out the Higgs field, we derived an EFT where the squared effective mass of the scalarization scalar matches that of Ref.~\cite{Eichhorn:2023iab}. However, in this scenario, vacuum GR solutions, such as the Kerr metric, no longer satisfy the equations of motion. Instead, scalarization would occur around stationary black hole solutions of a modified theory of the form $\mathcal{L} = R + k \GB^2$.

Future work could extend our analysis to the dynamical evolution of scalar perturbations around rotating Kerr black hole backgrounds, as well as to the study of spin-induced scalarization. Rotating solutions could be investigated using, for example, the numerical framework developed in Ref.~\cite{Fernandes:2022gde}. Additionally, it would be valuable to explore in greater detail whether vacuum solutions of the theory with the Higgs field are susceptible to scalarization, and if so, at which mass scales this occurs. A natural first step in this direction would involve studying static and spherically symmetric black holes within the EFT described by Eq.~\eqref{eq:higgs_EFT_ctescalar}, by employing both perturbation theory in the small-coupling limit and numerical techniques.

\acknowledgements
TS acknowledges partial support from the STFC Consolidated Grant nos. ST/V005596/1 and ST/X000672/1.
PF is funded by the Deutsche Forschungsgemeinschaft (DFG, German Research Foundation) under Germany’s Excellence Strategy EXC 2181/1 - 390900948 (the Heidelberg STRUCTURES Excellence Cluster).

\bibliography{bibnote.bib}
\end{document}